\documentclass[twocolumn,preprintnumbers,showpacs,pre,floatfix,amsmath,amssymb,notitlepage,superscriptaddress]{revtex4-1}
\usepackage{epsfig}
\usepackage{subfigure}
\usepackage{amsmath}
\usepackage{color}
\usepackage{amssymb}
\usepackage{setspace}
\usepackage{graphicx}% Include figure files
\usepackage{dcolumn}% Align table columns on decimal point
\usepackage{bm}% bold math
\usepackage{times}
\usepackage{enumerate}
\usepackage{footnote}

\input epsf

\graphicspath{{figures/}}

\begin{document}

\author{Per Sebastian Skardal}
\email{persebastian.skardal@trincoll.edu} 
\affiliation{Department of Mathematics, Trinity College, Hartford, CT 06106, USA}

\author{Alex Arenas}
%\email{persebastian.skardal@trincoll.edu} 
\affiliation{Department d'Enginyeria Inform\'{a}tica i Matem\'{a}tiques, Universitat Rovira i Virgili, 43007 Tarragona, Spain}

\title{Memory selection and information switching in oscillator networks with higher-order interactions}

%\date{\today}

\begin{abstract}
We study the dynamics of coupled oscillator networks with higher-order interactions and their ability to store information. In particular, the fixed points of these oscillator systems consist of two clusters of oscillators that become entrained at opposite phases, mapping easily to information more commonly represented by sequences of 0's and 1's. While $2^N$ such fixed point states exist in a system of $N$ oscillators, we find that a relatively small fraction of these are stable, as chosen by the network topology. To understand the memory selection of such oscillator networks, we derive a stability criterion to identify precisely which states are stable, i.e., which pieces of information are supported by the network. We also investigate the process by which the system can switch between different stable states when a random perturbation is applied that may force the system into the basin of attraction of another stable state.
\end{abstract}


\maketitle

\section{Introduction}\label{sec:01}

Collective behavior in ensembles of network-coupled dynamical units is an important area of research due to a wide range of both natural and engineered applications~\cite{Strogatz2003,Pikovsky2003}. Examples include cardiac pacemakers~\cite{Glass1988}, synthetic cell engineering~\cite{Prindle2012Nature}, and power grid dynamics~\cite{Rohden2012PRL,Skardal2015SciAdv}. Moving beyond such a system's ability to either remain incoherent or synchronize into a single entrained group, various combinations of dynamical and structural properties may give rise to more complicated synchronization patterns, for instance when oscillators form coexisting synchronized and incoherent groups known as chimera states~\cite{Panaggio2015Nonlin} or spontaneously partition themselves in different clusters that remain entrained within but do not synchronize with one another~\cite{Pecora2014NatComm}. Of particular interest here are patterns where phase oscillators form multiple entrained groups. In the illustrative case of two groups, which will be our focus in this work, oscillator systems can be used as models for information storage, treating oscillators that entrain to one or the other group as a 0 or 1, i.e., a bit, in a piece of information~\cite{Daido1996PRL,Ashwin2007SIAM,Skardal2011PRE,Komarov2013PRL}. Given the mappings between phase and integrate-and-fire oscillators~\cite{Politi2015}, understanding the dynamics that give rise to such patterns may shed light on cognitive function and memory~\cite{Arenas1994EPL,Vicente1996JPA,Hoppensteadt1999PRL,Fell2011Nature,Hipp2012Nature}.

In addition to the collective behavior seen in network dynamics that emerge from classical pair-wise interactions, a great deal of interest has recently been paid to higher-order interactions, i.e., interactions that take place between three or more units (and are fundamentally different than linear combinations of pair-wise interactions)~\cite{Salnikov2019EJP,Battiston2020PR,Carletti2020JPhys}. A large outstanding question in this direction is: What novel effects do such higher-order interactions have on macroscopic behavior? Phase reduction techniques have already shown that such interactions take place in generic oscillator systems~\cite{Ashwin2016PhysD,Leon2019PRE} and empirical data suggests that they may play an important role in brain dynamics~\cite{Petri2014Interface,Giusti2016JCN,Sizemore2018JCN}. In particular, three-way interactions interactions may describe correlations in neuronal spiking activity in the brain that provides a missing link between structure and function~\cite{Reimann2017}. Some results have been developed for treating synchronization of identical (possibly chaotic) oscillators~\cite{Mulas2020PRE,Gambuzza2020} and higher-order interactions between phase oscillators have been investigated in a handful of studies~\cite{Tanaka2011PRL,Komarov2015PRE,Bick2016Chaos,Millan2020PRL,Lucas2020}, but we demonstrated in recent work that such interactions naturally enable systems to store memory and information via the kinds of patterns described above~\cite{Skardal2019PRL,Xu2020PRR}.

In this paper we study the process by which memory is generated in oscillator systems with non-trivial network topologies and higher-order interactions. We find that, compared to the all-to-all coupled case studied in Refs.~\cite{Skardal2019PRL,Xu2020PRR}, when there is an underlying network topology that constrains the interactions the memory capacity of the system, quantified by the number of stable fixed point states the dynamics supports, is also constrained. Specifically, we identify precisely which oscillator states, i.e., pieces of information, are supported by a given network topology using a linear stability criterion. We then study the process by which the system switches from one piece of information to another as perturbations are applied to the different stable states and may move the system into the basin of attraction of a different state.

The remainder of this paper is organized as follows. In Sec.~\ref{sec:02} we present the model and discuss the connection between stable states and memory. In Sec.~\ref{sec:03} we present a stability analysis and explore memory capacity in oscillator networks with higher-order interactions with an illustrative example. In Sec.~\ref{sec:04} we investigate the process by which the system can move between different stable states under random perturbations, thereby switching between pieces of information. In Sec.~\ref{sec:05} we explore the effects of network structure and investigate the asymmetry of stable states. In Sec.~\ref{sec:06} we conclude with a discussion of our results.

\section{The 2-Simplex Oscillator Model}\label{sec:02}

We begin by consider an extension of the Kuramoto model with higher-order interactions, specifically three-way, or 2-simplex, interactions~\cite{Kuramoto1984}. The all-to-all coupled case was treated in detail in Refs.~\cite{Skardal2019PRL,Xu2020PRR}, but when placed on a nontrivial network topology the system is governed by the following equations of motion for $N$ oscillators,
\begin{align}
\dot{\theta}_i=\omega_i+\frac{K}{2}\sum_{j=1}^N\sum_{l=1}^NB_{ijl}\sin(\theta_j+\theta_l-2\theta_i),\label{eq:01}
\end{align}
where $\theta_i$ and $\omega_i$ are the phase and natural frequency, respectively, of oscillator $i$ and $K$ is the global coupling strength. Importantly, the coupling function is the sine of a linear combination of three oscillators, rather than two, and interactions are determined by the adjacency tensor $B$. We note that this coupling function is one of the two possibilities for 2-simplex interactions that arise from phase reduction. The other possibility, $\sin(2\theta_j-\theta_l-\theta_i)$, does not admit stable solutions with multiple clusters, and therefore does not support memory storage, so we focus on the coupling in Eq.~(\ref{eq:01}). For simplicity we assume that the entries of $B$ are unweighted and undirected, so that $B_{ijl}=1$ if a three-way interaction exists between oscillators $i$, $j$, and $l$ (and other wise $B_{ijl}=0$) and $B_{ijl}=B_{ilj}=B_{jil}=B_{jli}=B_{lij}=B_{lji}$. This last property essentially states that interactions are not directed and, with the factor of $2$ in the denominator of Eq.~(\ref{eq:01}), each three-way interactions is counted exactly once. In principle, the higher-order structure encoded in $B$ may be defined in different ways depending on the application. In other words, the entries of $B$ may or may not depend on the entries of an underlying network consisting of nodes and links (i.e., 0- and 1-simplexes) encoded in an adjacency matrix $A$. However, here we consider the case where higher-order interactions are in fact dictated by the adjacency matrix $A$ (which we assume is unweighted and undirected) so that $B_{ijl}=A_{ij}A_{jl}A_{li}$, i.e., a higher-order interaction exists between a group of three nodes if they are connected in a triangle.

The dynamics of Eq.~(\ref{eq:01}) naturally give rise to synchronized state patterns that entrain oscillators in two groups at opposite sides of the torus. To see this we introduce the set of local order parameters
\begin{align}
z_i = r_i e^{i\psi_i}=\frac{1}{2}\sum_{j=1}^N\sum_{l=1}^NB_{ijl}e^{i\theta_j}e^{i\theta_l},\label{eq:02}
\end{align}
which allows us to rewrite Eq.~(\ref{eq:01}) as 
\begin{align}
\dot{\theta}_i=\omega_i+Kr_i\sin(\psi_i-2\theta_i).\label{eq:03}
\end{align}
First, by entering the rotating reference frame $\theta\mapsto\theta+\overline{\omega}t$, where $\overline{\omega}=\frac{1}{N}\sum_{i=1}^N\omega_i$ is the mean natural frequency, we may set the mean frequency to zero. Next, assuming that the dynamics relax to a stationary state, oscillator $i$ becomes phase-locked if $|\omega_i|\le Kr_i$ (we will return to this condition later) and converges to one of two stable fixed points defined by 
\begin{align}
\theta_i^*=\phi_i~~~\text{or}~~~\phi_i+\pi.\label{eq:04}
\end{align}
where
\begin{align}
\phi_i^*=\frac{\psi_i+\arcsin\left(\frac{\omega_i}{Kr_i}\right)}{2}.\label{eq:05}
\end{align}
A suitable shift initial conditions allows us to center the mean phases $\psi_i$ about zero, and under typical conditions we expect that phase-locking oscillators will have $\psi_i\approx0$, which, combined with Eq.~(\ref{eq:04}), yields two entrained groups of phase-locked oscillators: one about $\theta=0$ and the other about $\theta=\pi$.

While this demonstrates that phase-locked oscillators tend to fall into one of two possible groups, the presence of drifting oscillators poses an issue in terms of interpretation. To this end, we consider the dynamics of Eq.~(\ref{eq:01}) in the regime where the coupling strength is sufficiently strong compared to the spread of frequencies, so that $K\gg|\omega_i|$ for all $i=1,\dots,N$. By considering the rescaled time $\tau=K t$ and approximating $\omega_i/K\approx0$, Eq.~(\ref{eq:01}) simplifies to
\begin{align}
\dot{\theta}_i=\frac{1}{2}\sum_{j=1}^N\sum_{l=1}^NB_{ijl}\sin(\theta_j+\theta_l-2\theta_i),\label{eq:06}
\end{align}
in which case all oscillators converge to either $0$ or $\pi$.

Before we move on to the analysis of these states, it's worthwhile to discuss their representation of information. In particular, a fixed point of a system of $N$ oscillators will be given by some $\bm{\theta}^*\in\mathbb{R}^N$, where each $\theta_i^*=0$ or $\pi$. This is easily interpretable as a string of bits: for instance, in a system of $N=5$ oscillators we may converge to the state $\bm{\theta}^*=[0,0,\pi,0,\pi]^T$, which one can be mapped to the sequence of bits $(0,0,1,0,1)$. This manner of memory in oscillator systems is not dissimilar to that observed in neural models consisting of Ising spin particles, best exemplified in the seminal 1988 papers by Gardner~\cite{Gardner1988JPA1} and Gardner and Derrida~\cite{Gardner1988JPA2}. In these models information bits, i.e., 0's and 1's, are equivalent not to oscillator phases, but Ising spins: pluses and minuses. The oscillator systems studied here can then be interpreted as a different model for capturing and studying a similar phenomenon.

\begin{figure*}[t]
\centering
\epsfig{file =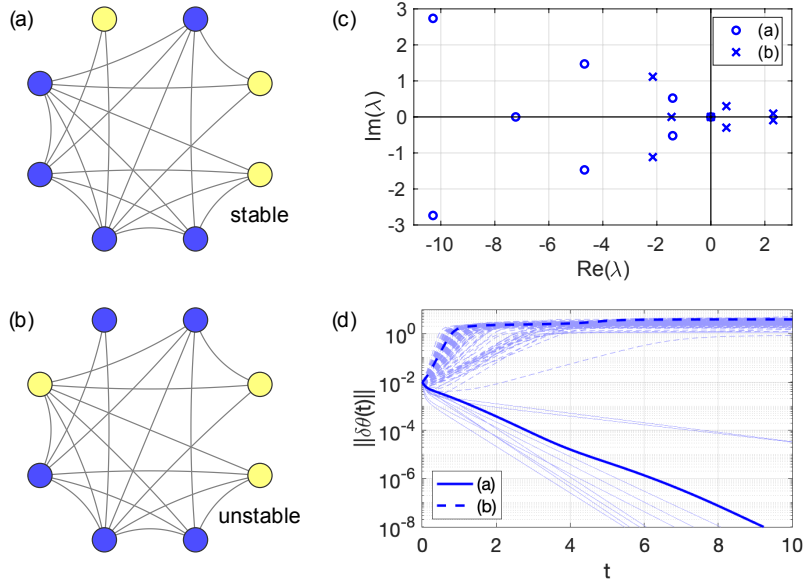, clip =,width=0.65\linewidth }
\caption{{\it Example: Stable and unstable states.} (a), (b) Two possible fixed points of the dynamics that are stable and unstable, respectively, on a network of size $N=8$. Oscillator values $\theta=0$ and $\pi$ are colored blue and yellow, respectively. (The top right oscillator is constrained to be $\theta_1^*=0$.) (c) Eigenvalue spectrum of the Jacobian $DF(\bm{\theta}^*)$ for states depicted in (a) and (b), plotted as circles and crosses, respectively. (d) The time series of the norm of perturbations $\bm{\delta\theta}(t)=\bm{\theta}(t)-\bm{\theta}^*$, initially set to $\|\bm{\delta\theta}(0)\|=10^{-2}$, for all $2^{N-1}=128$ possible states. Perturbations to states depicted in (a) and (b) are plotted as thick solid and dashed curves, respectively. Perturbations to all other possible states are plotted as thin curves, solid or dashed if they are predicted to be stable or unstable, respectively.} \label{fig1}
\end{figure*}

\section{Stability Analysis}\label{sec:03}

To understand the nature of fixed point states of the dynamics given by Eq.~(\ref{eq:06}), we begin by analyzing their asymptotic stability. Recall that we are interested in fixed points $\bm{\theta}^*\in\mathbb{R}^N$ with entries $\theta_i^*=0$ or $\pi$. First, $2^N$ such states exist. However, the dynamics are invariant under rotations, including rotation by $\pi$, i.e., $\theta\mapsto\theta+\pi$, so without any loss of generality we ``anchor'' $\theta_1^*=0$, leaving $2^{N-1}$ possible fixed point solutions. Then, given a fixed point $\bm{\theta}^*$, its stability is governed by the eigenvalues of the Jacobian matrix $DF(\bm{\theta}^*)$ whose entries are given by
\begin{widetext}
\begin{align}
DF_{ij}(\bm{\theta}^*)=\left\{\begin{array}{rl}-\sum_{k=1}^N\sum_{l=1}^NB_{ikl}\cos(\theta_k^*+\theta_l^*-2\theta_i^*) & \text{if }i=j,\\ \sum_{l=1}^NB_{ijl}\cos(\theta_j^*+\theta_l^*-2\theta_i^*) & \text{if }i\ne j.\end{array}\right.\label{eq:07}
\end{align}
\end{widetext}
Due to the rotational invariance of the dynamics, $DF(\bm{\theta}^*)$ is guaranteed to have one trivial eigenvalue $\lambda=0$, which can also be seen by noting that the rows of $DF(\bm{\theta}^*)$ all sum to zero, so $\bm{v}\propto\bm{1}$ is an eigenvector with eigenvalue $\lambda=0$. Therefore, if all other eigenvalue have strictly negative real part, i.e., are contained in the left-half complex plane, then $\bm{\theta}^*$ is stable, and otherwise it is unstable. This gives us the machinery to evaluate the stability of each of the $2^{N-1}$ fixed points described above.

\begin{figure*}[t]
\centering
\epsfig{file =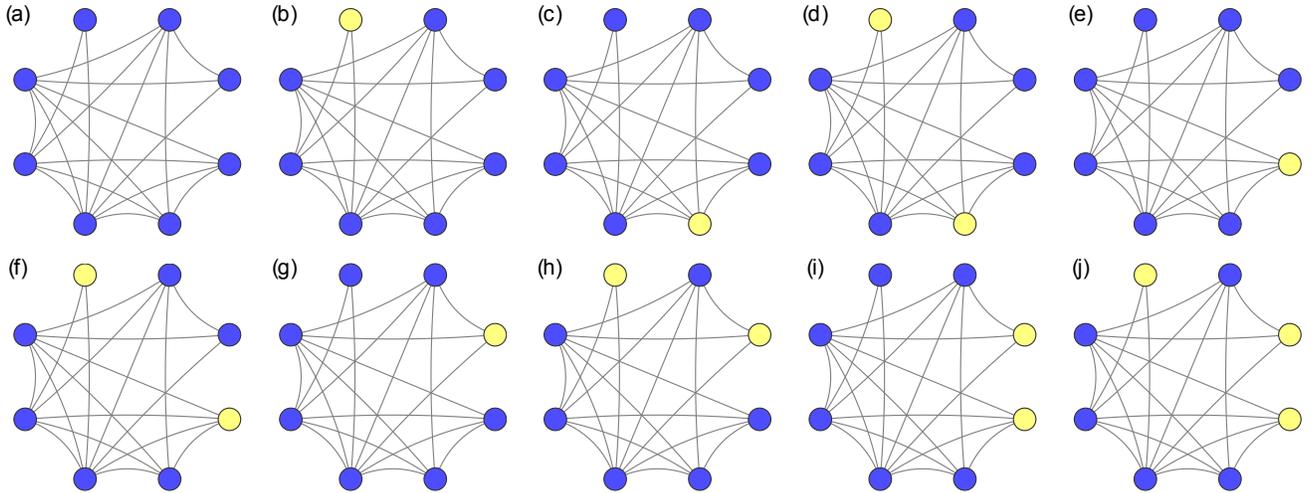, clip =,width=\linewidth }
\caption{{\it Stable states.} For the example network depicted in Fig.~\ref{fig1}, all $10$ stable states (out of a possible $2^{N-1}=128$ possible fixed points).} \label{fig2}
\end{figure*}

Before a more in-depth exploration of the stability of possible states we consider a small, illustrative example. Specifically, consider the network and two states depicted in Fig.~\ref{fig1} (a) and (b), where blue and yellow filled nodes correspond to states $\theta_i=0$ and $\pi$, respectively. (Here we index the nodes $i=1,2,\dots,N=8$ in the clockwise direction, starting from the top, so we constrain $\theta_1=0$.) For the given network structure (where all nodes are part of at least one connected triangle and therefore are part of at least one higher-order interaction) the state depicted in (a) is stable, while the state in (b) is unstable. To see this we plot in Fig.~\ref{fig1}(c) the eigenvalue spectra of the Jacobian matrices $DF(\bm{\theta}^*)$ for the two states (a) and (b) using circles and crosses, respectively. Not that in with exception to the trivial eigenvalue $\lambda=0$ for each, the full spectrum for (a) is contained in the left-half complex plane, whereas four eigenvalues for (b) are located in the right-half complex plane. To test this stability criterion we directly simulate the evolution of a small perturbation $\bm{\delta\theta}(t)=\bm{\theta}(t)-\bm{\theta}^*$ initially set at $\|\bm{\delta\theta}(0)\|=10^{-2}$ and plot the results in Fig.~\ref{fig1}(d). Time series for perturbations to states (a) and (b) are plotted as thick solid and dashed curves, respectively, showing that the perturbations exponentially decay and grow until saturation, respective. We also plot the time series for perturbations to all other states as solid or dashed curves based on the prediction from the stability criterion, getting $100\%$ agreement. 

Continuing with the example above, only a relatively small fraction (10 out of a possible $2^{N-1}=128$) of states are in fact stable. In Fig.~\ref{fig2} we plot these stable states. In terms of strings of bits (where $\theta=0$ and $\pi$ correspond to $0$ and $1$) these states are precisely
\begin{align}
&(0,0,0,0,0,0,0,0),~~(0,0,0,0,0,0,0,1),\nonumber\\
&(0,0,0,1,0,0,0,0),~~(0,0,0,1,0,0,0,1),\nonumber\\
&(0,0,1,0,0,0,0,0),~~(0,0,1,0,0,0,0,1),\label{eq:08}\\
&(0,1,0,0,0,0,0,0),~~(0,1,0,0,0,0,0,1),\nonumber\\
&(0,1,1,0,0,0,0,0),~~(0,1,1,0,0,0,0,1)\nonumber
\end{align}
Inspecting these ten states more closely, we find two important properties. First, a number of the stable states appear to be combinations of one another in the sense that oscillators with $\theta^*_i=\pi$ in different stable states often combined to create another stable state. For example, the state depicted in Fig.~\ref{fig2}(d) is a combination of those in Figs.~\ref{fig2}(b) and (c). This does not always happen, however, as states depicted in Figs.~\ref{fig2}(c) and (e) do not combine to form a new stable state. More specifically, it also points to the role of node $i=8$ (the location of $\theta_i=\pi$ in the state at the top of the right column) which has no effect on the stability of a given state. To see this, note that any stable state with $\theta_8=0$ is still stable when $\theta_8$ is changed to $0$, and vice versa. It is worth noting that node $i=8$ is the most poorly connected node in the network is the sense that is has only one three-way interaction.

The second important property that this example illustrates is a certain asymmetry in the collection of stable states. Note first, that of the ten total stable states, one has zero $\theta^*=\pi$ entries, four have one $\pi$ entry, four have two $\pi$ entry, and one has three $\pi$ entries. However, of all $2^{N-1}=128$ possible fixed points, one has zero $\pi$ entries, seven have one $\pi$ entry, 21 have two $\pi$ entries, 35 have three $\pi$ entries, 35 have four $\pi$ entries, and so on. In particular, the majority of all fixed point states have a roughly even distribution of $0$ and $\pi$ entries, but when we restrict our attention to only stable states we find a strong asymmetry as they tend to have an uneven distribution of $0$ and $\pi$ entries.

\begin{figure}[t]
\centering
\epsfig{file =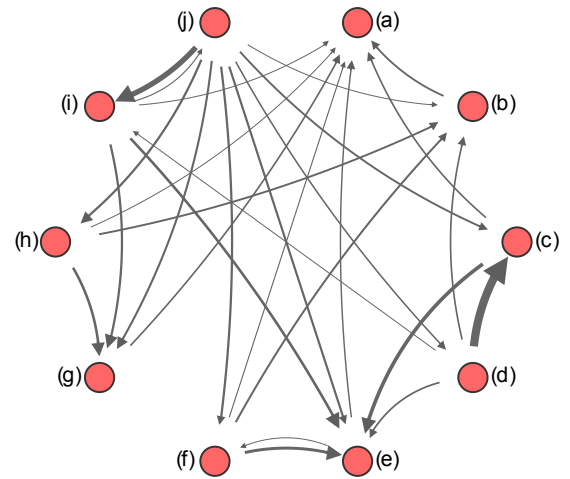, clip =,width=0.9\linewidth }
\caption{{\it Switching.} Illustration of the transition network between states (a)--(j) illustrated in Fig.~\ref{fig2}. Each directed link is drawn with width proportional to the rate at which the system switches from one state to another under a random perturbation of size $\|\bm{\delta\theta}\|=1.55$. (Self-links, describing rates at which perturbations do not cause a transition, are not illustrated.)} \label{fig3}
\end{figure}

\section{Switching}\label{sec:04}

Lastly, with the stability analysis above as a tool for identifying stable states of the system, i.e., supported pieces of information, we investigate the process by which the system may switch between different pieces of memory, i.e., stable states. What we describe here is not quite a homoclinic network~\cite{Ashwin2013PhysD,Bick2018PRE} but is in many senses similar. Sticking with our example network illustrated above, none of the 10 stable states depicted in Fig.~\ref{fig2} are connected with a heteroclinic orbit since they are all asymptotically stable. Instead, we again consider perturbations $\bm{\delta\theta}$ to these stable states $\bm{\theta}^*$, but now we allow perturbations to not be arbitrarily small, letting $\|\bm{\delta\theta}\|=\mathcal{O}(1)$, so that the perturbed state $\bm{\theta}^*+\bm{\delta\theta}$ may fall outside of the basin of attraction of $\bm{\theta}^*$ and eventually converge to another stable state.

More precisely, we consider the following setup. For each of the 10 stable states (a)--(j) depicted in Fig.~\ref{fig2}, we introduce $10^3$ random perturbations of size $\|\bm{\delta\theta}\|=1.55$. Then, for each of these perturbations to the states $j$ we find the fraction that eventually converge to each other state $i$, and let this fraction populate the entry $\pi_{ij}$ of the transition matrix $\Pi$. Thus, each entry $\pi_{ij}$ describes the rate at which perturbations of this size cause a switch in states $j\to i$. This transition network is by nature directed and weighted, and is illustrated in Fig.~\ref{fig3}, letting more strongly (weakly) weighted links be thicker (thinner). We have neglected to illustrate self-links, i.e., entries $\pi_{ii}$ that describe rates at which the perturbation does not result in a switch in state. It should be noted that the choice $\|\bm{\delta\theta}\|=1.55$ is important; choosing perturbations too small eventually leads to no transitions from one state to another, while choosing perturbations too large eventually leads to much denser networks whose links reflect only the size of the basins of attraction of each state. The choice used here lies in between these two extremes. In this example, for instance, we see that a few state pairs have particularly strong rates of transitions, i.e., (d)$\to$(c) and (j)$\to$(i). This stems from the similarity of the states themselves (see Fig.~\ref{fig2} where it can be seen that these states differ only by a single oscillator). Switching between such similar states appears to exist between other pairs as well, e.g., (b)$\to$(a), (f)$\to$(e), and (h)$\to$(g), but the strength of the transition is variable. Moreover, the uniform state (a) (where $\bm{\theta}^*=\bm{0}$) is a sink where the system cannot escape from. This property is not particularly surprising for reasonable perturbations since state (a) is the most linearly stable in the sense that the largest nontrivial eigenvalue is the most negative compared to those for other states.

\begin{figure}[t]
\centering
\epsfig{file =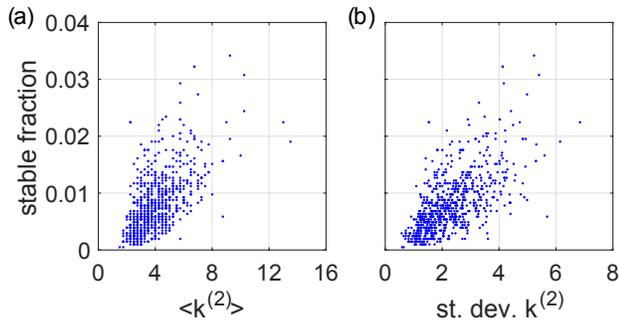, clip =,width=1.0\linewidth }
\caption{{\it Effect of network structure.} Illustration of the transition network between states (a)--(j) illustrated in Fig.~\ref{fig2}. Each directed link is drawn with width proportional to the rate at which the system switches from one state to another under a random perturbation of size $\|\bm{\delta\theta}\|=1.55$. (Self-links, describing rates at which perturbations do not cause a transition, are not illustrated.)} \label{fig4}
\end{figure}

\section{Network Structure and State Asymmetry}\label{sec:05}

Before concluding we investigate two aspects of the stability properties of the system. We begin by probing the effect of network structure on the overall stability of different states. Keeping in mind the combinatorial complexity of the number of possible states to probe (recall that for a network of $N$ oscillators we have $2^{N-1}$ states to examine) we maintain a relatively small network size, specifically $N=12$, but study an ensemble of $800$ soft spherical networks built using the hyperbolic graph generator~\cite{Aldecoa2015CPC}. In particular, we use a small but finite temperature of $T=1$ guaranteeing that, while some long-range links exist, the networks are significantly clustered. (We throw out any networks that are not connected by its triangles.) Moreover, while we use a target mean 1-simplex degree of $\langle k^{(1)}\rangle=4$ we note that the mean 2-simplex degree, $\langle k^{(2)}\rangle$, where $k_i^{(2)}=\frac{1}{2}\sum_{j,l=1}^NA_{ij}A_{jl}A_{li}$, as well as its standard deviation, vary significantly from network to network.

We begin by investigating the overall stability of states by, for each network, calculating the fraction of all states that are stable. In Fig.~\ref{fig4}(a) and (b) we plot the fraction of stable states for each network vs the corresponding mean and standard deviation of the 2-simplex degree. It's clear to see that the fraction of stable states that a network admits has a positive correlation with both the mean and standard deviation of the 2-simplex degree.

\begin{figure}[t]
\centering
\epsfig{file =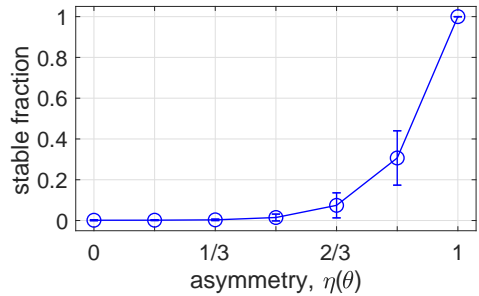, clip =,width=0.8\linewidth }
\caption{{\it Asymmetry of stable states.} Illustration of the transition network between states (a)--(j) illustrated in Fig.~\ref{fig2}. Each directed link is drawn with width proportional to the rate at which the system switches from one state to another under a random perturbation of size $\|\bm{\delta\theta}\|=1.55$. (Self-links, describing rates at which perturbations do not cause a transition, are not illustrated.)} \label{fig5}
\end{figure}

We also investigate more closely the nature of the states that end up being stable. Recall that from the example depicted in Fig.~\ref{fig2} we observed a noticeable asymmetry. We now examine this by defining an asymmetry identifier for each state $\bm{\theta}^*$ as
\begin{align}
\eta(\bm{\theta}^*)=\frac{\sum_{i=1}^N\chi_{\theta_i=0}-\sum_{i=1}^N\chi_{\theta_i=\pi}}{N},\label{eq:09}
\end{align}
where $\chi_{\theta_i^*=0} = 1$ if $\theta_i^*=0$ and $\chi_{\theta_i^*=0} = 0$ otherwise (and similarly, $\chi_{\theta_i^*=\pi} = 1$ if $\theta_i^*=\pi$ and $\chi_{\theta_i^*=\pi} = 0$ otherwise). In particular, $\eta(\bm{\theta}^*)=1$ or $0$ in the cases that all $\theta_i^*$ share the same value or are even split between $0$ and $\pi$. Using this asymmetry identifier to organize the entire ensemble of possible states, we then evaluate, for all states $\bm{\theta}^*$ that share an asymmetry identifier value $\eta(\bm{\theta}^*)$ the fraction of states that are stable. In Fig.~\ref{fig5} we plot, as a function of the asymmetry identifier $\eta(\bm{\theta}^*)$ the fraction of stable states for states sharing the value $\eta(\bm{\theta}^*)$ averaged over the ensemble of networks used above. (Error bars indicate one standard deviation above and below the mean.) We observe that the trend observed in our example depicted in Fig.~\ref{fig2} persists, i.e., states that are more asymmetric are more likely to be stable.

\section{Discussion}\label{sec:06}

In this paper we have studied the process by which oscillator networks with higher-order interactions can select memory and information, often represented as sequences of bits, by stabilizing fixed points consisting of two groups of clustered oscillators at opposite phases. While oscillator networks have long been used as models for encoding and storing such information~\cite{Hoppensteadt1999PRL}, such systems have often required fine-tuning and engineering of, e.g., individual coupling strengths, to attain a given state. However, the discovery that higher-order interactions naturally leads to a a wide variety of such states provides a more natural formalism for information storage without such fine-tuning~\cite{Skardal2019PRL,Xu2020PRR}.

Here we have presented a stability analysis for determining precisely which of the $2^N$ potential states are stable. This selection depends on the underlying network topology that encodes the higher-order interactions between oscillators. Moreover, we have also used our stability analysis to study the process by which the oscillator network can switch between different stable states, as a perturbation can be applied to force the system into a basin of attraction for another stable state.

\bibliographystyle{plain}

\end{document}